\documentclass[aps,prl,twocolumn,letterpaper,superscriptaddress]{revtex4-1}
\usepackage{graphicx,amsmath,amssymb,amsfonts,latexsym,color,dcolumn,bm,epsfig,subfigure}
\usepackage[plainpages=false,hyperfootnotes=false,colorlinks=false]{hyperref}

\renewcommand{\imath}[0]{\mathrm{i}}

\begin{document}

\title{Nonadditive Enhancement of Nonequilibrium Atom-Surface Interactions} 

\author{D. Reiche}\email[Corresponding author.\\]{reiche@physik.hu-berlin.de}
\affiliation{Humboldt-Universit\"at zu Berlin, Institut f\"ur Physik, AG 
			Theoretische Optik \& Photonik, 12489 Berlin, Germany}
\affiliation{Max-Born-Institut, 12489 Berlin, Germany}

\author{K. Busch}
\affiliation{Humboldt-Universit\"at zu Berlin, Institut f\"ur Physik, AG
			Theoretische Optik \& Photonik, 12489 Berlin, Germany}
\affiliation{Max-Born-Institut, 12489 Berlin, Germany}

\author{F. Intravaia}
\affiliation{Humboldt-Universit\"at zu Berlin, Institut f\"ur Physik, AG
			Theoretische Optik \& Photonik, 12489 Berlin, Germany}

\newcommand{\mathbfh}[1]{\hat{\mathbf{#1}}}

\begin{abstract} 
The motion-induced drag force acting on a particle moving parallel to an arrangement of $N$ objects is analyzed. 
Particular focus is placed on the nonequilibrium statistics of the interaction and on the interplay between the system's geometry and the different dissipative 
processes occurring in realistic setups.
We show that the drag force can exhibit a markedly nonadditive enhancement
with respect to the corresponding additive approximation. 
The specific case of a planar cavity -- a relevant configuration 
for many experiments -- is calculated, showing an enhancement of about one order of magnitude.
This and similar configurations are of significant potential interest for future measurements that aim to detect the drag force.

\end{abstract}

\maketitle


According to quantum electrodynamics, there is no free space in the classical sense of an empty vacuum
\cite{Dirac27}. 
Instead, free space is filled with zero-point fluctuations and the state of this quantum vacuum is not 
unique: It is strongly influenced by any material body and the motion of the observer. 
Zero-point fluctuations 
induce forces that act on any form of matter. 
Mostly quantum in nature, such forces can 
display quite unintuitive characteristics. 
Prominent representatives are van der Waals and Casimir-Polder forces 
\cite{Intravaia11}. 
Very interestingly, when acting on a particle in the vicinity of macroscopic bodies, these interactions are found to be \emph{nonadditive} and dependent on the system's geometry \cite{Milonni92a}. 
This property can be utilized to tailor the interaction and both experimental and theoretical investigations
have demonstrated nonadditive corrections of up to 50\% 
\cite{Chan08,Intravaia13,Hartmann17,Garrett18}.
If the system is driven out of equilibrium, fluctuation-induced forces can have additional
intriguing aspects due to their inherent connection to the system's underlying statistics. In this 
case, to the best of our knowledge, nonadditive behavior has been investigated for temperature gradients 	
\cite{Antezza06} or external optical fields \cite{Fuchs18a} only. 
In this manuscript, we show that \textit{mechanical} nonequilibrium situations can allow for a strong nonadditive enhancement of about one order of magnitude or larger, considerably improving the chances for an experimental demonstration.

The technological progress of recent years has allowed to control the motion of particles in highly
confined spaces such as the inside of cavities or optical fibers.
Typical physical examples include atoms \cite{Ritsch13,Epple14}, large molecules \cite{Hornberger12}, dielectric or metallic nanoparticles \cite{Bykov15},
and nitrogen-vacancy centers (NV-centers) in nanodiamonds \cite{Tisler13,Schell14,Farias20}.
For a moving particle, the interaction with the (quantum) electromagnetic fluctuations in close proximity to an object leads to a force that acts parallel to the object's surface, which -- at 
zero temperature -- 
is referred to as quantum friction 
\cite{Dedkov02a,Volokitin07,Scheel09,Maghrebi13,Jentschura15,Viotti19,Farias20}. 
Recent work has highlighted the relevance of nonequilibrium physics in the context of this 
phenomenon as well as the importance of the materials' dissipative properties in characterizing its strength and its dependence on the particle's velocity~
\cite{Intravaia14,Intravaia16,Reiche17,Oelschlager18}.

\begin{figure}
  \centering
    \includegraphics[width=0.34\textwidth]{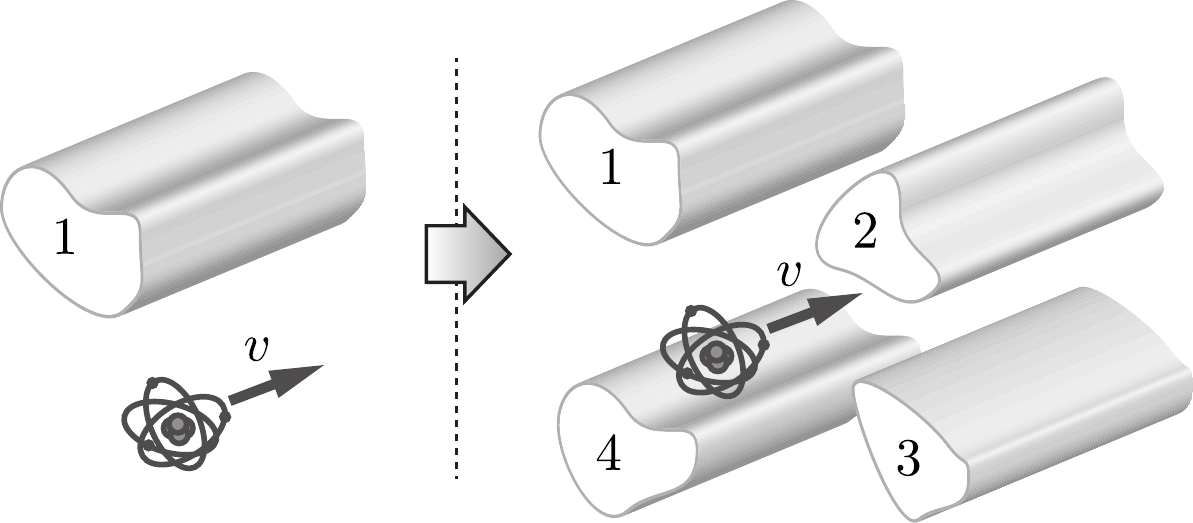}
    \vspace{-0.3cm}
    \caption{
	 A particle moving parallel to one or
an arrangement of $N$ different, translationally invariant objects. \label{fig:generic-setup}}
		  \vspace{-0.3cm}
\end{figure}

To investigate how dissipation and non-additivity combine in quantum friction, we consider a neutral, polarizable particle nonrelativistically moving parallel to 
an arrangement of $N$ objects. 
This arrangement is translation invariant along the direction of motion
(Fig. \ref{fig:generic-setup}). 
The particle is described using its electric dipole operator $\mathbfh{d}(t)$ and each body is comprised of an isotropic, reciprocal, linear and homogeneous material. 
Further, we assume that, at late times, the system reaches a nonequilibrium steady state (NESS) and moves at constant speed $v$ \cite{Intravaia15,Intravaia16a}. This presupposes the existence of an unspecified external mechanism that balances the drag force. 
Proceeding similarly to Ref. \cite{Intravaia16a}, one can show that quantum friction acts opposite to the direction of motion and its strength is given by
\begin{multline}
   F = - 2\, \mathrm{Tr}\int_{0}^{\infty} \mathrm{d}\omega
	         \int \frac{\mathrm{d} q}{2\pi} \, q ~\underline{S}^{\sf T}(-\omega^{-}_{q},v)
			                                      \underline{G}_{\Im}(q,\mathbf{R}_{a}, \omega)   .
\label{totalForce}
\end{multline} 
The superscript ``$\sf T$'' gives the transpose of a matrix, $q$ denotes the component of the radiation's wave vector parallel to the direction of motion, $\mathbf{R}_{a}$ the transversal position of the particle and $\omega^{\pm}_{q} = \omega\pm q v$ is the Doppler-shifted frequency. 
Physically, Eq.~\eqref{totalForce} can be regarded as
being the result of the total momentum per unit of time transferred to the particle 
during the absorption and emission of excitations extracted from vacuum 
\cite{Maghrebi13,Intravaia15,Intravaia16b}.
The processes are described by two quantities: The Green tensor $\underline{G}$ with $\underline{G}_{\Im}=[\underline{G}-\underline{G}^{\dagger}]/(2\imath)$, connected to the $N$-bodies' electromagnetic response, and the power spectrum 
$\underline{S}(\omega,v)$, determining the statistical properties of the particle's internal dynamics.

In contrast to the Casimir-Polder force, Eq.~\eqref{totalForce} has been mainly evaluated for the case of one single planar surface.
When more objects are present, the additive approximation suggests that $F\approx F_{\rm add}=\sum_{i=1}^{N} F_{i}$, where $F_{i}$ is the force occurring when only the $i$th body is present. 
This indicates the existence of specific configurations where the force can be enhanced by a factor $\sim N$ with respect to a single body result.
Formally, $F_{i}$ can be calculated from Eq.~\eqref{totalForce} by replacing $\underline{G}$ with $\underline{G}_{i}$, describing the $i$th body alone. Intuitively, one expects then that the additive description works if $\underline{G}\approx \sum_{i=1}^{N}\underline{G}_{i}$.
Clearly, this ignores the mutual interactions between the objects that are responsible for some nonadditive behavior observed in the equilibrium case. 
In mechanical nonequilibrium configurations, however, intriguing additional nonadditive features emerge.

The behavior of our system is strongly connected to the expression for $\underline{G}$ which is in general rather involved.
Still, some general remarks useful for our analysis are possible. Given that a Hermitian matrix can be decomposed as the sum of a symmetric and an anti-symmetric term \cite{dennis03}, we have that
\begin{equation}
\label{split}
\underline{G}_{\Im}(q,\mathbf{R}_a,\omega)=\underline{\Sigma}(q,\mathbf{R}_a,\omega)+\mathbf{s}_{\perp}(q,\mathbf{R}_a,\omega)\cdot\mathbf{\underline{L}},
\end{equation}
where $\underline{L}_{i}=-\imath\epsilon_{ijk}$ is the generator of rotations around the $i$-axis.
Due to the passivity of the materials comprising the bodies, the matrix $\underline{\Sigma}$ is real, symmetric, positive semidefinite for $\omega\ge0$ and even in $q$.
The real vector $\mathbf{s}_{\perp}$ is odd in $q$ and, for symmetry reasons, orthogonal to the direction of invariance.
It can be related to a spin-dependent part of the electromagnetic density of states \cite{Intravaia19a,Mandel95}, including the so-called spin-momentum locking of light \cite{Bliokh15,Lodahl17} (see Refs. \cite{OShea13,Sayrin15,Gong18} for recent experiments).

Turning to the power spectrum, its form is deeply connected to the system's dissipative dynamics. 
In quantum electrodynamic systems, one can essentially distinguish two different physical damping mechanisms. The first is \textit{intrinsic} dissipation, arising from a large number of degrees 
of freedom inside the particle itself. 
Examples are ro-vibrational modes and/or cross-state electronic interactions in molecules \cite{Reitz19}, electron-electron or electron-phonon scattering in metallic nanoparticles \cite{Bass90,Rubio-Lopez18a}, and vibrations or deformations within diamond lattice-embedded NV-centers \cite{Behunin16}. The second source of dissipation is \textit{radiation-induced} damping, which originates 
from the interaction of the system with the (quantized) electromagnetic field: 
Light and matter degrees of freedom mix (dressing) to give rise to hybrid polaritonic states. 
The dressing is responsible for frequency shifts and line-broadening in the particle's spectrum, as also recently investigated in the field of molecular polaritonics \cite{Sanvitto16,Feist18}.
In general, both damping mechanisms are interlaced but, depending on the system, the role played by one can be more relevant than the other. 

Contrary to previous approaches treating quantum friction, in order to describe both these processes on the same footing, we take the particle's electric dipole to linearly interact with both the electromagnetic field \emph{and} 
with a bath accounting for internal losses. 
Instead of diagonalizing the system's (very large) Hamiltonian, we focus on the stationary solutions of its equations of motion (Heisenberg picture) and combine them with linear response theory \cite{Kubo57}.
Specifically, in the limit where the dipole's fluctuating dynamics can be modeled in terms of an isotropic Drude-Lorentz oscillator \cite{Note1}, we can write
\begin{multline}\label{oscillator}
	\ddot{\mathbfh{d}}(t)+\epsilon_{0}\omega_{a}^{2}
	\int\mathrm{d}t_{1}\;\underline{\mu}(t-t_{1})
			\dot{\mathbfh{d}}(t_{1})+\omega_a^2\mathbfh{d}(t)\\
			=\alpha_0\omega_a^2\left[\mathbfh{f}_{0}(t)+\mathbfh{E}(\mathbf{r}_a(t),t)\right],
\end{multline}
where  $\mathbf{r}_a(t)$ is the particle's trajectory, $\omega_a$ its internal electronic
transition frequency and $\alpha_0$ its static polarizability
\cite{Note2}. 
If we, for the time being, disregard the electric field $\mathbfh{E}$, Eq.~\eqref{oscillator} 
is the three-dimensional generalization of the so-called quantum Langevin equation  
\cite{Ford87a,Ford88a}: 
The term $\mathbfh{f}_{0}$ is the bath's Langevin force operator and is related to the free evolution of the internal degrees of freedom. It is connected to the response kernel $\underline{\mu}(\tau)=\langle (\imath/\hbar) \theta(\tau)[\mathbfh{f}_{0}(\tau),\mathbfh{f}_{0}(0)]\rangle$ 
\cite{Kubo57,Note3}
[or its Fourier transform $\underline{\mu}(\omega)$]
via the fluctuation-dissipation theorem \cite{Kubo66,Callen51},
\begin{equation}
\langle \mathbfh{f}_{0}(\omega)\mathbfh{f}_{0}(\omega')\rangle= 4\pi \hbar\, \theta(\omega)\, \alpha_{0}^{-1}\epsilon_{0} \omega\, \underline{\mu}_{\Re}(\omega)\delta(\omega+\omega'),
\end{equation}
where $\underline{\mu}_{\Re}=[\underline{\mu}+\underline{\mu}^{\dag}]/2$, $\theta(x)$ is the Heaviside function and the brackets denote the quantum average over the initial state of the system (assumed to be factorized).
As any response function \cite{Dressel02}, the expression for $\underline{\mu}(\omega)$ can be rather involved and it depends on the parameters 
defining the particle's internal degrees of freedom (e.g. their energy spectrum). Its expression, however, is also strictly 
constrained by thermodynamic considerations \cite{Ford88a}, requiring that $\underline{\mu}_{\Re}$ is positive semidefinite.
Note that, despite the oscillator's coupling constant is a scalar ($\alpha_{0}$), we allow for an 
anisotropic internal dissipation through the tensorial form of $\underline{\mu}$.
It is reasonable to assume the statistical independence of the dissipative mechanisms and require
that $\langle\mathbfh{E}_0(\mathbf{r},t)\mathbfh{f}_{0}(t)\rangle=0$ \cite{Note4}, where
$\mathbfh{E}_0$ describes the stationary quantum electromagnetic field \emph{without} the particle \cite{Intravaia16a}. 
In this case the power spectrum tensor reads
\begin{subequations}
\label{FDR}
\begin{gather}
\underline{S}(\omega,v)
	=
	\frac{\hbar}{\pi}
	\underline{\alpha}(\omega,v)
	\underline{\mathcal{D}}(\omega,v)
	\underline{\alpha}^{\dag}(\omega,v),\\
\underline{\alpha}(\omega,v)
	=
	\underline{\alpha}_{\mu}(\omega)
	\left[1-\int
	\frac{\mathrm{d}q}{2\pi}
	\underline{G}(q,\mathbf{R}_a,\omega^{+}_{q})\underline{\alpha}_{\mu}(\omega)
	\right]^{-1},
\end{gather}
\end{subequations}
where $\underline{\alpha}(\omega,v)$ and 
$\underline{\alpha}_{\mu}(\omega)=\alpha_0
	\left[
	1-\omega^2/\omega_a^2-\imath \epsilon_{0} \omega\underline{\mu}(\omega)
	\right]^{-1}$ 
are, respectively, the velocity dependent and the \emph{intrinsically}-damped polarizabilities. 
In the nonequilibrium fluctuation relation presented in Eqs.~\eqref{FDR}, the dissipation 
kernel $\underline{\mathcal{D}}(\omega,v)$ results from the two-time correlator of the 
(quantum) noise terms associated with the different dissipation mechanisms, i.e.
\begin{equation}
\label{dissKernel}	
\underline{\mathcal{D}}(\omega,v)
	=\frac{\omega \epsilon_{0}\theta(\omega)}{\alpha_0}\underline{\mu}_{\Re}(\omega)
	+\int \frac{\mathrm{d}q}{2\pi}\theta(\omega^{+}_{q})
	~\underline{G}_{\Im}(q,\mathbf{R}_a,\omega_{q}^+).
\end{equation}
As a consequence of the assumption that the vacuum field and the Langevin force are
uncorrelated, the two dissipation channels 
add linearly in $\underline{\mathcal{D}}(\omega,v)$. Notice that this result is not
trivial, because of the system's NESS. 
Nevertheless, Eqs. \eqref{FDR} show that in general the field and the bath interlace in their contribution to the 
dipole's statistical dynamics and the dressed polarizability.

The $\theta(\omega)$ appearing in Eq. \eqref{dissKernel} essentially limits the 
frequency integration in Eq. \eqref{totalForce} to $0<\omega\lesssim q v$, indicating that the
dominant contribution to friction arises from the evanescent sector \cite{Oelschlager18}.
This diminishes the relevance of resonant effects that are connected with multiple interferences of propagating waves.
Generally, we have that $|q|\lesssim 1/\lambda$, where $\lambda$ is a length-scale related to the system's geometry, material and $|\mathbf{R}_{a}|$ \cite{tai94,SuppMat}. 
Inserting typical values shows that quantum friction is essentially a low-frequency phenomenon.
As a consequence, to leading order coupling $\alpha_0$, we can approximate 
$\underline{S}(\omega,v)
 \approx
 (\hbar \alpha_{0}^{2}/\pi) \underline{\mathcal{D}}(\omega,v)$ in Eq.~\eqref{totalForce}.
Accordingly, the dissipative mechanisms decouple 
and the force can be written as $F\approx F^{\rm int}+F^{\rm rad}$, where $F^{\rm int}$ 
is connected to the particle's intrinsic dissipation while $F^{\rm rad}$ to radiation
damping. We have \cite{Note5}
\begin{subequations}
\label{forces}
\begin{align}
\label{internal}
   F^{\rm int} =
 -  \frac{\hbar \alpha_{0}\epsilon_{0}}{\pi}v^{3}
	         &\int_{0}^{\infty} \frac{\mathrm{d}q}{2\pi} \, 
	         \frac{q^{4}}{3}
    \mathrm{Tr} \left[\underline{\mu}^{\sf T}_{\Re}(0)\underline{\Sigma}'(q, \mathbf{R}_{a},0)\right],	  
\end{align}
\begin{align}
\label{radiative}
   F^{\rm rad} =
   - \frac{\hbar \alpha_{0}^{2}}{\pi}v^{3}
	       &  \int \frac{\mathrm{d}q}{2\pi} 
	         \int \frac{\mathrm{d}\tilde{q}}{2\pi}
	         \, 
	       \frac{(\tilde{q}+q)^{4}}{12}
	           \nonumber  \\
	           & \times \left\{\mathrm{Tr} \left[
	\underline{\Sigma}'(\tilde{q},\mathbf{R}_{a}, 0) \underline{\Sigma}'(q, \mathbf{R}_{a}, 0)\right]\right.
\nonumber\\	
	           &\quad \left.-2 
	~\mathbf{s}'_{\perp}(\tilde{q},\mathbf{R}_{a}, 0)\cdot \mathbf{s}'_{\perp}(q, \mathbf{R}_{a}, 0)
	\right\},
\end{align}
\end{subequations}
where the prime indicates the derivative with respect to frequency, which we assume to be nonzero at $\omega=0$ \cite{SuppMat}.

While both components are negative and hence counteract the motion, their expressions are strikingly different. 
$F^{\rm int}$ is linear in the Green tensor and, in agreement with our additive intuition, an increase in the number of objects can lead to an enhancement $\sim N$ of the force. This can occur for example when $N$ identical objects are placed around the particle trajectory.
Clearly, the linear growth with $N$ is at some point limited by different factors such as size and proximity of the objects. These also include nonadditive contributions, which are expected to arise from the quasi-electrostatic interactions between the bodies and from the corresponding frequency shifts of the polaritonic excitations living on each of them \cite{Intravaia05,Intravaia07,Moeferdt18}. Their impact on the force becomes stronger the closer the bodies are.

The component of quantum friction associated with radiative damping, $F^{\rm rad}$, 
features a much more intriguing behavior and contains the main result of this paper. 
When contrasted with Eq.~\eqref{internal}, Eq.~\eqref{radiative} reveals 
the rather distinct physical processes that underlie the radiative dissipation channel: While $F^{\rm int}$ results from an interplay of the internal dissipation with the 
electromagnetic environment and vanishes in the limit $\underline{\mu}\to 0$, $F^{\rm rad}$ 
is induced by the backaction of the field  
onto the particle, which persists even in the limit of vanishing internal 
damping. As a consequence,
unlike $F^{\rm int}$, $F^{\rm rad}$ is quadratic in the Green tensor, 
indicating that increasing the number of objects can be responsible of a nonadditive 
enhancement $\propto N^{2}$ with respect to the single object configuration 
\cite{NoteMaster}.
Perhaps more surprising is that
from Eq. \eqref{radiative} we can see that $F^{\rm rad}$ consists of two distinct contributions, containing either $\underline{\Sigma}$ or $\mathbf{s}_{\perp}$. 
Usually, they tend to almost compensate each other \cite{SuppMat}, making the force smaller. For an atom moving near a single surface, this leads to a reduction in strength of about 70\% \cite{Intravaia19a}.
Physically, the contribution associated with $\mathbf{s}_{\perp}$ stems from 
the coupling between the particle's translational and rotational degrees of freedom and
involves a selective exchange of angular momentum with the field. 
However, if several bodies are placed around the particle, for symmetry reasons, this process can be inhibited, effectively uncoupling translational and rotational motion. Specifically, the vector $\mathbf{s}_{\perp}$ vanishes if $\mathbf{R}_{a}$ is located on the symmetry axis of an axis-symmetric configuration.
In other words, the enhancement with respect to the single-object configuration is rather of the form $\sim \phi N^{2}$, where $\phi>1$ is the typical factor due to the suppression of this mechanism ($\phi \sim 3.5$ for a planar surface \cite{Intravaia19a}).

\begin{figure}
  \centering
    \includegraphics[width=0.40\textwidth]{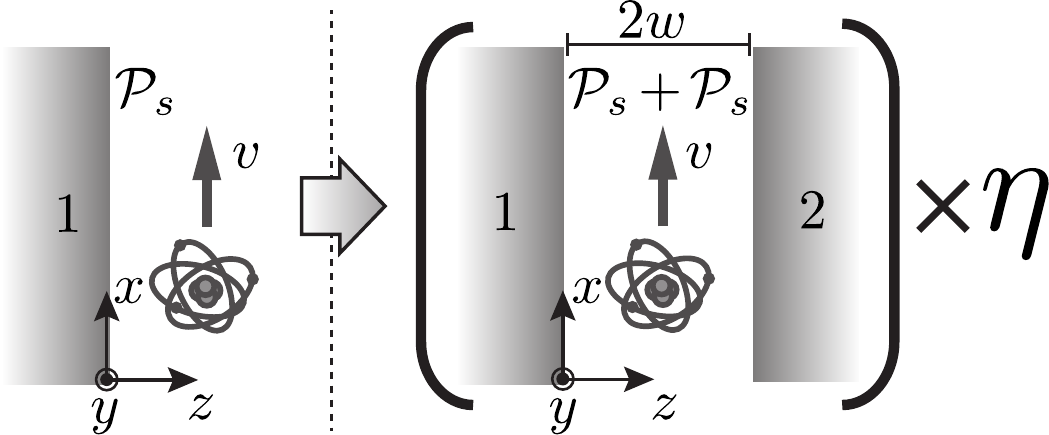}
    \vspace{-0.3cm}
    \caption{\label{fig:setup}
	\textit{Left}: A particle moving parallel to a single interface. $\mathcal{P}_s$ is related to the electromagnetic response of the interface \cite{SuppMat}.
    \textit{Right}: A particle moving inside a planar cavity of width $2w$ 
		parallel to the (potentially distinct) material surfaces \cite{SuppMat}. 
    The factor $\eta$ describes the non-additivity of the frictional 
		force: 
		$\eta = 1$ corresponds to the additive description. }
		  \vspace{-0.3cm}
\end{figure}

For further insights, it is interesting to consider the example of a polarizable particle moving within a planar cavity of width $2w$ (Fig. \ref{fig:setup}).
For such a simple configuration the expression for the Green tensor is available \cite{SuppMat}.
Without loss of generality, we choose the $x$-axis as the direction of motion  
and further assume that the $xy$-plane coincides with one plane of the cavity. The 
$z$-axis points into the cavity such that for the particle's position we have $z_a\in(0,2w)$. 
For simplicity, we assume first that the 
plates are identical ($r_{1,2}=r$). 
In the quasistatic regime ($\omega\rightarrow0$),
cavity resonances become unimportant and wave vectors are limited by $1/\lambda\sim\text{max}\left(z_a^{-1},[2w-z_a]^{-1}\right)$. 
The maximal deviation from the additive expression occurs for the maximum distance from the surfaces,  i.e., for $z_{a}=w$.  
Quantitatively, we can introduce a non-additivity factor $\eta^{\rm int}=F^{\rm int}/F_{\rm add}^{\rm int}$, where $F_{\rm add}$ gives the corresponding expressions of the naive addition of two separate surfaces. 
For $\underline{\mu}\equiv\mu$ and identical (spatially local)
plates, we obtain
\begin{equation}
\label{F-Int-constr}
1<\eta^{\rm int}(z_{a})\le \frac{1}{15} \left(\frac{\pi}{2}\right)^{6},
\end{equation}
which corresponds to a correction of about $0.14\%$ over the entire range of the particle's
positions $z_{a}$ within the cavity \cite{Note6}.
Notably, within the range of validity of our description \cite{SuppMat}, the bounds of the previous 
relation are independent of the size of the cavity.
They only depend on the static value of the reflection coefficient and saturate for 
$r(\omega=0)=1$. 
For two plates made from different materials, the largest non-additivity 
is achieved in a position closer to the one plate that exhibits lower dissipation. 
Depending on the difference in material properties, we can also exceed the upper 
bound of Eq. (\ref{F-Int-constr}). Surprisingly, friction is also enhanced with respect to a single plane even by introducing a second surface made from a perfectly conducting material, despite this interface does not generate any friction by itself \cite{Intravaia16a}. 
This can be understood either in terms of an effectively larger number of image dipoles interacting
with the one dissipative surface \cite{Intravaia16a} or equivalently through a shift of the surface plasmon-polariton frequency induced by the boundary conditions of the perfectly conducting material. 
Remarkably, for anisotropic internal dissipation, $F_{\rm add}^{\rm int}$ might both over- and underestimate the value 
of $F^{\rm int}$. 
Again, the largest 
deviation for identical plates is observed at the center of the cavity.  
For $\mu_{zz}=0\not=\mu_{xx}=\mu_{yy}$ and 
$\mu_{zz}\not=0=\mu_{xx}=\mu_{yy}$,  
non-additivity amounts to roughly $\mp 2\%$, respectively \cite{SuppMat}.

A comparison of $F^{\rm rad}$ for 
identical plates with its additive approximation gives instead
\begin{equation}
\label{maxAtom}
1<\eta^{\rm rad}(z_{a})\le \frac{13249}{56700}\left(\frac{\pi}{2}\right)^{8}\approx 8.66,
\end{equation}
which reveals a nonadditive enhancement of about one order or magnitude and, accordingly to the previous general analysis, a force which is 17 times larger than the single-plane result. As in the case of internal dissipation, 
the largest deviation is observed for $z_{a}=w$ when $r(\omega=0)=1$. 
\begin{figure}[!t]
  \centering
    \includegraphics[width=0.45
 \textwidth]{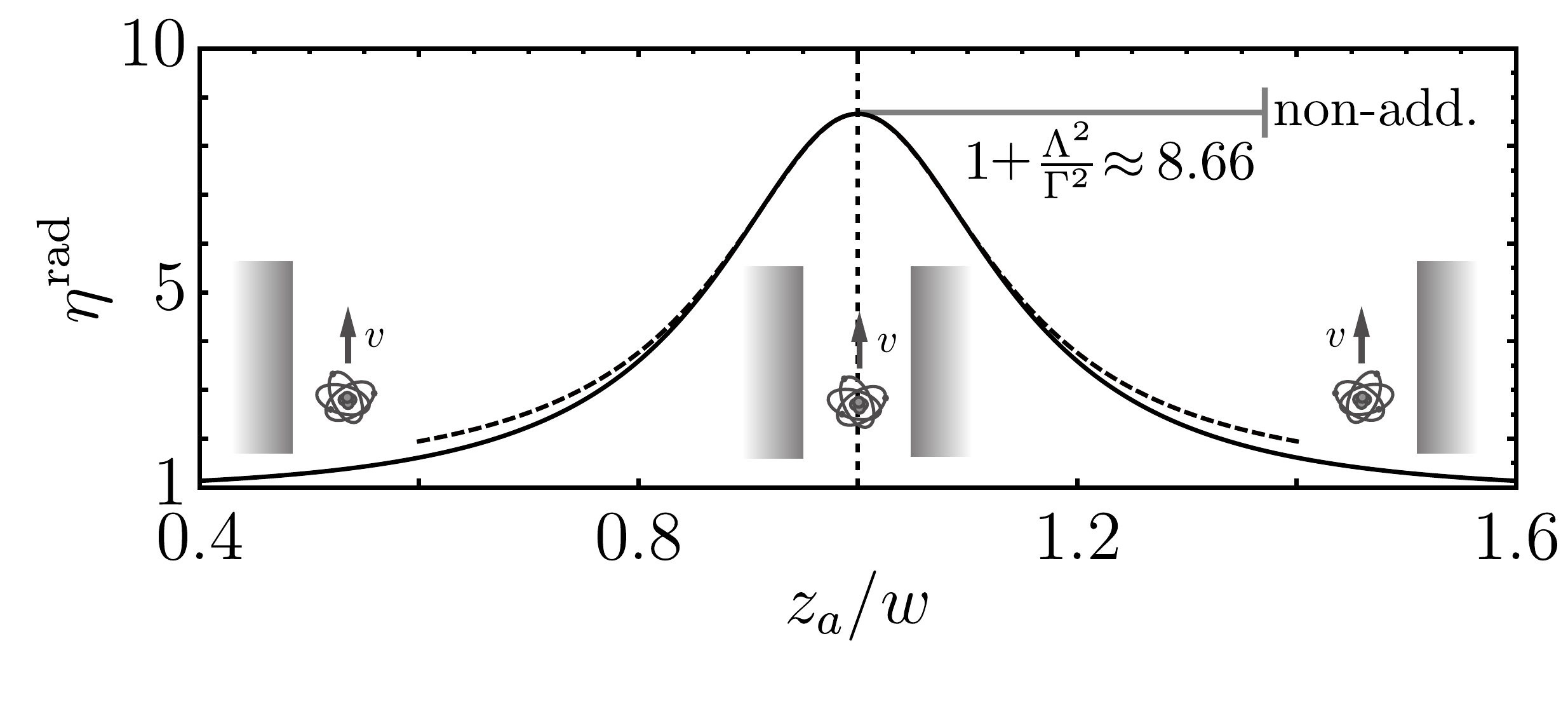}
    \vspace{-.7cm}
    \caption{Nonadditive enhancement of the frictional force in the limit $\underline{\mu}\to 0$ (e.g. for atoms) 
		         as a function of the particle's position inside the cavity [see Eq.~\eqref{maxAtom}]. 
						 We have used $r(0)=1$. The dashed line is the approximation  
						 in Eq.~\eqref{LorentzAppr}.}
 \label{fig:f}
\end{figure}
We represent $\eta^{\rm rad}(z_{a})$ in Fig.~\ref{fig:f} and approximate its Lorentzian-like shape by
\begin{equation}
\eta^{\rm rad}(z_{a})\approx 1+\frac{\Lambda^{2}}{\left(1-z_{a}/w\right)^{2}+\Gamma^{2}},
\label{LorentzAppr}
\end{equation}
where $\Lambda\approx 0.42$ and $\Gamma\approx 0.15$.
As for $F^{\rm int}$, we can exceed the above bound by considering 
different materials for the cavity's plates.

It is important to notice that the nonadditive enhancement described above can be combined with other mechanisms affecting the strength of the force. 
These range from the use of specific materials \cite{Kalusniak14} to more intriguing phenomena connected with nonlocality \cite{Reiche17,Reiche19}. 
Even the structure of each single object can be relevant \cite{Oelschlager18}:  Eqs.~\eqref{forces} share indeed relevant features with the Purcell factor \cite{Novotny06} and surface-enhanced spectroscopy \cite{Moskovits85}.
While the frequency value restrains any resonant amplification, an enhancement can still arise from a tighter field confinement and smaller mode volume \cite{kristensen19a}.

We would like to remark that the two-plate geometry is also close to some already existing experimental setups. 
One of the most prominent is the diffraction of an atomic or molecular beam on a grating \cite{Hornberger12,Fein19}, where high-velocity particles ($\sim$ km/s \cite{Brand19}) are impinging on slits having a width in the range of a few tens of nanometers \cite{Brand15}. 
Due to the contactless interaction between the particle and the internal wall of the slit, 
the wave function describing the quantum-mechanical dynamics of the beam acquires a phase 
that can be visible within the interference pattern that forms behind the grating \cite{Perreault05a,Lepoutre11}. 
Also, in microfabricated collimators, atoms can already fly at the speed 
of sound in narrow capillary-like structures over lengths of millimeters \cite{Li19}.
Alternatively, one might consider atom-interferometric setups, where one arm of the interferometer 
is led through a waveguide: 
Here, a combination of lasers can also provide the driving force and a stabilizing 
potential 
\cite{Bykov15}. 

Our results stress that dissipation and geometry nontrivially interlace in quantum friction, highlighting 
how the fundamental properties of the material-modified quantum vacuum behave in this situation.
A careful design and/or a structural engineering 
of the system can have a severe impact on fluctuation-induced forces in mechanical nonequilibrium,
significantly increasing an usually weak effect. 
Already, a four-plate configuration points to a possible enhancement factor of $\phi N^2\sim 56$.
A broader study of this and similar setups is therefore promising, prompting towards higher 
chances of success for a future experimental demonstration.


\paragraph{Acknowledgments}
We thank Ch. Egerland, M. Oelschl\"ager, B. Leykauf and Ch. Brand for stimulating discussions. 
We acknowledge support from the Deutsche Forschungsgemeinschaft (DFG, German Research Foundation) -- Project-ID 182087777 -- SFB 951.
F.I. further acknowledges financial support from the DFG through the DIP program (Grants
FO 703/2-1 and SCHM 1049/7-1). 
D.R. is grateful for support from the German-American Fulbright Commission (Doktorandenprogramm).
%


\clearpage

\section*{\Large Supplemental Material}

\subsection{On the sign of the two components of $F^{\rm rad}$}

In the main text, Eq. (7b) for $F^{\rm rad}$ features two different contributions. The first is associated with the real positive semi-definite (for $\omega\ge 0$) matrix $\underline{\Sigma}(q,\mathbf{R}_a,\omega)$ and the second with the real vector $\mathbf{s}_{\perp}(q,\mathbf{R}_a,\omega)$.
While the former is a even function of $q$, the second is an odd function of the same variable. 
The complete integrand of Eq. (7b) contains a positive function multiplied by a difference of two terms. 
The subtraction arises from $\mathrm{Tr}[\underline{L}^{\sf T}_{i}\underline{L}_{j}]=-2\delta_{ij}$, which corresponds to a selection rule in the exchange of angular momentum between the atom and the field.
We have that the term containing $\mathrm{Tr} \left[
	\underline{\Sigma}'(\tilde{q},\mathbf{R}_{a}, 0) \underline{\Sigma}'(q, \mathbf{R}_{a}, 0)\right]$ is positive because it is related to the trace of the product of two positive semi-definite matrices. 	
Determining the sign of contribution due to the integral containing $\mathbf{s}'_{\perp}(\tilde{q},\mathbf{R}_{a}, 0)\cdot \mathbf{s}'_{\perp}(q, \mathbf{R}_{a}, 0)=\sum_{i}s'_{i}(\tilde{q}, \mathbf{R}_{a}, 0)s'_{i}(q, \mathbf{R}_{a}, 0)$ requires more care. 
Expanding the $(\tilde{q}+q)^{4}$, we obtain
\begin{multline}
\label{spin_contribution}
\sum_{i} \int\mathrm{d}q\,\mathrm{d}\tilde{q} \;(\tilde{q}+q)^{4}s'_{i}(\tilde{q}, \mathbf{R}_{a}, 0)s'_{i}(q, \mathbf{R}_{a}, 0)
\\
 =32\sum_{i} \int_{0}^{\infty}\mathrm{d}q \;q^{3}s'_{i}(q, \mathbf{R}_{a}, 0)\int_{0}^{\infty}\mathrm{d}q \;q \; s'_{i}(q, \mathbf{R}_{a}, 0).
\end{multline}
Therefore the two-dimensional integral in the first line of the previous expression is clearly positive if either $s'_{i}(q, \mathbf{R}_{a}, 0)>0$ or $s'_{i}(q, \mathbf{R}_{a}, 0)<0$ for all $q>0$, i.e. if the function does not oscillate as a function of $q>0$. 
This is typically the case since we are working in the evanescent region and using common materials (see also below). 
It is interesting to mention, however, that in general the expression in Eq. \eqref{spin_contribution} can still be positive even if $s'_{i}(q, \mathbf{R}_{a}, 0)$ oscillates.

\subsection*{Planar cavity}

As described in the main text, as a specific example of our general description, we consider a polarizable particle within a planar cavity of width $2w$ 
moving with non-relativistic speed parallel to the cavity's material surfaces. The component of the wave vector parallel to the surface in indicated by $\mathbf{p}=(p_{x},p_{y})$ ($p=|\mathbf{p}|$) and, since the motion is along the $x$-direction, $q \equiv p_{x}$.
For $w$ smaller than the plasma wavelength $\lambda_p$ of the materials that comprise the plates 
(e.g. $\lambda_p\sim 150$~nm for metals \cite{Barchiesi14} and/or up to $\sim 1\mu$m for doped-semiconductors \cite{Kalusniak14}), the force is the strongest.
In this limit (near-field region), $\underline{G}$ is dominated by the TM-polarized reflection coefficients 
$r_{1,2}(p,\omega)$ of the cavity interfaces, while the 
contribution of the TE-polarization can be neglected. The relevant contribution is provided by the scattered part of the Green tensor \cite{dedkov03,tomas95}
\begin{multline}\label{Green}
\underline{G}(p_{x},z_a,\omega)
\approx \int\frac{\mathrm{d}p_{y}}{2\pi} \left\{\frac{p}{2\epsilon_0}	
	\mathcal{P}_{+}(p,z_a,\omega)	
	\underline{\Pi}\right.
	\\		
	\left.-\frac{p}{2\epsilon_0}\mathcal{R}(p,\omega)
	\underline{\rm M}
	\cdot
	\underline{\Pi}
	-\frac{p_x}{2\epsilon_0}
	\mathcal{P}_{-}(p,z_a,\omega)
	\underline{L}_y\right\}
	,
\end{multline}
where $\epsilon_0$ is the vacuum permittivity and
\begin{subequations}\label{PR}
\begin{gather}	
	\label{R}
\mathcal{R}
	=2\frac{r_{1}r_{2}e^{-4pw}}{1-r_{1}r_{2}e^{-4pw}},
\\
\label{P}
\mathcal{P}_{\pm}
	=\frac{e^{-2pw}\left[r_{1}e^{2p \left(w-z_a\right)}
	\pm r_{2}e^{-2p\left(w-z_a\right)}\right]}{1-r_{1}r_{2}e^{-4pw}}.
\end{gather}
\end{subequations}
As indicated by the characteristic denominator, each of the above terms 
includes Fabry-Perot reflections associated with cavity systems. Further, 
we have defined 
$\underline{\Pi}=\text{diag}[p_x^2/p^2,p_y^2/p^2,1]$ 
and 
$\underline{\rm M}=\text{diag}[1,1,-1]$, 
where the latter matrix describes the mirror reflection at the $xy$-plane. 
The Green tensor for a single surface is recovered from Eq.~\eqref{Green} by 
setting $\mathcal{R}=0$ and 
$\mathcal{P}_{\pm}\to\mathcal{P}_{\rm s}= r \exp[-2pz_{a}]$.
When considering $\underline{G}_{\Im}=[\underline{G}-\underline{G}^{\dagger}]/(2\imath)$, Eq. \eqref{Green} indicates that $\underline{\Sigma}$ is diagonal and that
$\mathbf{s}_{\perp}\equiv (0,s_{y},0)$.

Further, we assume that at low frequencies the imaginary parts of the reflection 
coefficients scale linearly with frequency (Ohmic response, valid for most materials) and obtain 
to leading order in velocity 
\begin{subequations}
\label{twoforces}
\begin{align} \label{np}
&F^{\rm int}
	=-\alpha_0v^3\frac{\hbar}{12\pi}		
	\\\nonumber
	&\qquad\times
	\int\frac{\mathrm{d}^2\mathbf{p}}{(2\pi)^2}
	~ p p_x^4
	\left(\mathcal{P}_{+I}^{\prime}	
	\mathrm{Tr}\left[\underline{\mu}\cdot\underline{\Pi}\right]
	-
	\mathcal{R}_{I}^{\prime}
	\mathrm{Tr}\left[\underline{\mu}\cdot\underline{\rm M}\cdot
	\underline{\Pi}
	\right]\right),
	\\
&F^{\rm rad}	\label{forceatom}
	=
	-\alpha_0^2v^3\frac{\hbar}{\pi}
	\int\frac{\mathrm{d}^2\mathbf{p}}
			 {(2\pi)^2}
	\frac{\mathrm{d}^2\tilde{\mathbf{p}}}{(2\pi)^2}
	\frac{p\tilde{p}}{(2\epsilon_0)^2} 
	\\\nonumber
	&\times
	\left(	
	\left[
	\frac{p_x^4}{6}+\frac{p_x^2\tilde{p}_x^2}{2}
	\right]
	\left\{
	\left(
	\mathcal{P}_{+I}^{\prime}\mathcal{\tilde{P}}_{+I}^{\prime}
	+\mathcal{R}'_I\mathcal{\tilde{R}}'_I 
	\right)
	\mathrm{Tr} 
	\left[\underline{\Pi}\cdot\underline{\tilde{\Pi}}\right]
	\right.\right.
	\\\nonumber
	&
	\qquad\qquad\qquad\quad~\left.
	-\left(
	\mathcal{P}_{+I}^{\prime}\mathcal{\tilde{R}}'_I
	+\mathcal{R}'_I\mathcal{\tilde{P}}_{+I}^{\prime}
	 \right)
	\mathrm{Tr} 
	\left[
	\underline{\Pi}\cdot \underline{\rm M}\cdot\underline{\tilde{\Pi}}	
	\right]
	\right\}
	\\\nonumber
	&\quad\left.
	+\frac{p_x\tilde{p_x}}{p\tilde{p}}
	\left[\frac{p_x^3\tilde{p}_x}{2}+\frac{p_x\tilde{p}_x^3}{6}\right]	
	\mathcal{P}_{-I}^{\prime}\mathcal{\tilde{P}}_{-I}^{\prime}	
	\mathrm{Tr}\left[\underline{L}_y^{\sf T}\underline{L}_y\right]
	\right).
\end{align}
\end{subequations}
Equations \eqref{twoforces} are the equivalent of Eqs.~(7) in the main text 
for the specific case of a cavity.
Here, for the sake of readability, we have dropped the integrand's functional 
dependencies and all quantities are evaluated at $\omega=0$; the prime indicates 
the derivative with respect to frequency, the subscript ``$I$'' stands for an 
expression's imaginary part and the tilde indicates a dependence on $\tilde{p}$ 
instead of $p$. 
Explicitly, we have for example
$\mathcal{\tilde{P}}_{\pm I}^{\prime}
 =
 \mathrm{Im}[\partial_{\omega}\mathcal{P}_{\pm}(\tilde{p},z_a,\omega)]_{\vert \omega=0}$.
In the following, we consider spatially local and Ohmic material characteristics
so that at low frequencies $r_{I}\approx 2\epsilon_{0}\rho \omega$, where $\rho$ 
is a positive constant connected with the dissipation in the surface's material. More 
accurate descriptions, including e.g. spatial dispersion (where $\rho\equiv\rho(p)$  
\cite{Reiche17,Reiche19}), 
are most likely enhancing the effects described below.

We start by analyzing Eq. \eqref{np}. Interestingly, in the limit of an isotropic 
bath, 
$\underline{\mu}$ effectively becomes a scalar ($\underline{\mu}\to\mu$) and only the function $\mathcal{P}_{+}$ 
appears in Eq.~\eqref{np}. 
The term related with $\mathcal{R}$ identically vanishes, since in this case 
$\mathrm{Tr}[\underline{\rm M}\cdot\underline{\Pi}]=0$.
This behavior is connected to the isotropy of the static polarizability 
and would be modified as soon as static anisotropy ($\alpha_{0}\to\underline{\alpha}_{0}$) or higher orders in 
$\alpha_{0}$ are considered. 
Neglecting $\mathcal{R}$ and the denominator in Eq.~\eqref{P}, we can write 
\begin{equation}
\mathcal{P}_{+}(z_{a})\approx\mathcal{P}_{\rm s}(z_{a})+\mathcal{P}_{\rm s}(2w-z_{a}) 
\end{equation}
and therefore for the force $F^{\rm int}(z_{a})$ we have 
\begin{equation}
F_{\rm s}^{\rm int}(z_{a})+F_{\rm s}^{\rm int}(2w-z_{a})\equiv F_{\rm add}^{\rm int}(z_{a}),
\end{equation}
i.e. the sum of the two single-surface contributions with
\begin{equation}
F_{\rm s}^{\rm int}(z_{a})
	=-\frac{15}{(2\pi)^{2} }
	\hbar\alpha_0\epsilon_0
		 \frac{\left[5\mu_{xx}+\mu_{yy}+6\mu_{zz}\right]\rho v^3}
		 {(2z_a)^{7}}.
\end{equation}
In the quasi-static regime ($\omega\rightarrow0$),
cavity resonances become unimportant.
The multiple interference term responsible for the denominator in Eq.~\eqref{P} is 
relevant for small values of $p$ ($p\ll 1/w$) only. 
The dominant contribution to the 
total recoil momentum absorbed by the particle is given by wave vectors 
$p\lesssim \text{max}\left(z_a^{-1},[2w-z_a]^{-1}\right)$ and therefore the maximal deviation 
from the additive expression occurs 
for $z_{a}=w$.

For anisotropic dissipation, the term in Eq.~\eqref{np} proportional 
to $\mathcal{P}_{+I}^{\prime}$ is modified and the term containing $\mathcal{R}_{I}^{\prime}$ 
introduces a \emph{distance-independent} non-additive contribution. 
Remarkably, the sign of the 
contribution due to $\mathcal{R}_{I}^{\prime}$ can vary. The trace operator in Eq.~\eqref{np} selects only the 
diagonal part of $\underline{\mu}$ and for $\mu_{zz}>\mu_{xx},\mu_{yy}$ 
this term tends to increase the frictional 
force, while in the opposite case the drag is reduced.
As a result, $F_{\rm add}^{\rm int}$ might both over- and underestimate the value 
of $F^{\rm int}$. 
As for isotropic internal dissipation, the largest 
deviation for identical plates is observed at the center of the cavity.  

The quadratic structure of Eq.~\eqref{forceatom} directly points to a non-additive 
behavior of $F^{\rm rad}$ in response to a cavity-induced change of the electromagnetic density of states (emDOS). 
Importantly, however, as pointed out in the main text,
due to interferences and the participation of the rotational degrees of freedom, 
the non-additive correction goes beyond a simple quadratic enhancement. 
For clarity, we analyze again the case where at 
small frequencies $r_{1,2}\approx r(\omega=0)+2\mathrm{i}\epsilon_0\rho\omega$.
As above, we define the additive approximation as  \cite{Intravaia19a}
\begin{subequations}
\begin{gather}
F_{\rm add}^{\rm rad}(z_{a})
 \equiv
 F_{\rm s}^{\rm rad}(z_{a})+F_{\rm s}^{\rm rad}(2w-z_{a}), 
 \\
F_{\rm s}^{\rm rad}(z_{a})=-\frac{18\hbar}{\pi^{3}}\alpha_0^2\rho^2\frac{v^3}{(2z_a)^{10}} .
\end{gather}
\end{subequations}
We first consider the second and third line of Eq.~\eqref{forceatom}, involving the diagonal 
part of the Green tensor and being connected with the matrix $\underline{\Sigma}$. At the center of the cavity, the term containing only the $\mathcal{P}^{\prime}_{+I}$
function is responsible for an enhancement factor of about two with respect to the additive 
expression. 
The terms proportional to $\mathcal{R}^{\prime}_{I}$ are related to the anisotropy of the electromagnetically induced 
damping in the dissipation kernel [Eq.~(6) of the main text]. They are non-existent in the additive 
expression. 
The second term in the second line of Eq.~\eqref{forceatom} does not depend on the particle's 
position since it arises from constructive interference in the cavity's emDOS. 
Even more interesting is the last (fourth) line of Eq.~\eqref{forceatom}, which is connected with the vector $\mathbf{s}_{\perp}$. 
As discussed in the main text, in the single-plate case, this contribution tends to decrease the frictional force and, 
for a spatially local material, leads to a relative reduction of about 70\% 
\cite{Intravaia19a}.
For two identical parallel plates, however, this term vanishes at the center of the cavity ($\mathcal{P}_{-}=0$ 
for $z_{a}=w$) and starts to be significant only when $z_{a}$ describes a position close 
to one of the surfaces. Physically speaking, we have that the presence of the second surface 
tends to inhibit the net exchange of angular momentum between the particle and the electromagnetic 
field (the corresponding part of the emDOS
vanishes at $z_{a}=w$). 
These effects combined give rise to the non-addivite enhancement reported in the main text. 
In particular the coefficient $\eta^{\rm rad}(z_{a})$ has a symmetric Lorentzian-like shape [see Eq.~(10)], whose 
effective parameters $\Lambda$ and $\Gamma$ were found by
expanding Eq.~\eqref{forceatom} around $z_a\sim w$.
%
%
%
%
%


\begin{thebibliography}{10}
\bibitem{NoteMaster}
This is not to be confused with the photon emission rate of densely packed excited atoms scaling quadratically with their number (superradiance) \cite{gross82,sinha18}.
\end{thebibliography}
\end{document}